\documentclass[
   ,final            
  ]
  {aipproc}

\usepackage{amssymb}
\usepackage{amsmath}

\layoutstyle{6x9}

\begin{document}

\title{The Cosmological Energy Density of Neutrinos from Oscillation Measurements}

\author{Kevork Abazajian}{ address={Theoretical Division, MS B285, Los Alamos
  National Laboratory, Los Alamos, New Mexico 87545} }

\begin{abstract}
The emerging structure of the neutrino mass matrix, when combined with the
primordial element abundances, places the most stringent constraint on the
flavor asymmetries in the cosmological neutrino background and therefore its
energy density.  I review the mechanism of synchronized neutrino oscillations
in the an early universe with degenerate (asymmetric) neutrino and antineutrino
densities and the implications of refined measurements of neutrino parameters.
\end{abstract}

\maketitle


\section{Introduction}
The dawn of the era of precision cosmology has come with the observations of
anisotropies in the cosmic microwave background (CMB) with the Wilkinson
Microwave Anisotropy Probe (WMAP) over the whole sky to better than fundamental
uncertainty over a wide range in anisotropy scale~\cite{wmap}. Combined with
the three-dimensional galaxy distribution of the Sloan Digital Sky
Survey~\cite{Tegmark:2003uf}, a consistent picture has emerged for the standard
concordance cosmology:  a universe dominated by dark matter and dark energy
with structure growing from nearly scale-invariant adiabatic Gaussian density
perturbations.  In the simplest models, WMAP and SDSS measure the cosmological
matter density to nearly 10\%~\cite{sdss}.

Given the success of the standard concordance cosmology, it is tempting to
assume that the density of all cosmological matter and radiation components of
the universe are known to great precision.  However, the neutrino density,
often simply assumed to be fixed to its standard model value, is actually only
known to factors of its own magnitude when using the WMAP data
alone~\cite{nnucmb}.

One can hope to do better with primordial nucleosynthesis.  During primordial
nucleosynthesis, the nucleon beta-equilibrium weak interaction rates are
sensitive to the electron neutrino and antineutrino densities. The cosmic
expansion rate depends on the overall neutrino density, which sets when nuclear
reactions freeze-out.  These two effects can compensate each other and can
produce primordial element abundances for deuterium, helium and lithium that
are consistent with their observed abundances, as long as the nucleon density
is increased to allow the nuclear rates to keep up with the required increased
expansion rate~\cite{wagoner}.  The non-zero neutrino chemical potentials (or
degeneracy parameters) of this model led to its description as degenerate big
bang nucleosynthesis (DBBN). Since the nucleon (baryon) density is
independently constrained by the CMB, the magnitude of deviations from non-zero
neutrino chemical potentials was appreciably constrained from the original DBBN
models, but still allowed neutrino densities over twice that of the standard
value~\cite{newdbbn}.

With the emergence of the mass and mixing spectrum of the active neutrino
flavors, particularly in the large to maximal mixing angles of the solar and
atmospheric neutrino oscillation solutions, it was proposed that the mixing
could lead to the equilibration of neutrino asymmetries prior to nucleosynthesis
in the studies of Refs.~\cite{fuller,lunardini}.

The first attempt to solve the full evolution equations for the active neutrino
system using was performed numerically by Dolgov, Hansen, Pastor, Petcov,
Raffelt, and Semikoz~\cite{dhpprs}, who found that the maximal mixing solution
of the atmospheric results and large mixing angle solution of the solar
neutrino problem invariably led to a near equalization of neutrino asymmetries
between flavors.  Therefore, DBBN, which required a large disparity between
electron and muon or tau neutrino densities, would not be viable in a universe
with the observed neutrino mass and mixing matrix.  Analytic insight into the
flavor asymmetries' equalization and a quantification of changes within the
range of mixing parameters was studied by Wong~\cite{yyy}, and Abazajian, Beacom
and Bell~\cite{abb}.  The constraint imposed by the resulting equalizing
transformations excludes DBBN and requires neutrino densities to be within
$\sim$3\% of the standard value.  Therefore, any non-standard cosmic radiation
energy density must come from a more exotic phenomenon than photons and
neutrinos .


\section{Synchronized Oscillations}
In an elegant paper, Pastor, Raffelt \& Semikoz~\cite{prs} showed that the
synchronization mechanism, initially studied in Refs.~\cite{Samuel} can be
framed in the representation of synchronized dipoles precessing in a magnetic
field, with the orientation of the dipole representing the flavor content.

The system of mixed neutrinos in a dense, scattering, self-refractive
environment must be handled in a density matrix formalism.  The two-flavor
neutrino density matrix is 
\begin{eqnarray}
\rho(p) = \left( \begin{array}{cc}
\rho_{\alpha\alpha} & \rho_{\alpha \beta} \\
 \rho_{\beta \alpha} & \rho_{\beta \beta}
\end{array} \right)
=  \frac{1}{2} \left[ P_0(p)
+ \boldsymbol\sigma\cdot{\mathbf P}(p)\right],
\end{eqnarray}
where ${\mathbf P}(p)$ as the neutrino's ``polarization'' vector, which can be
represented as an individual ``magnetic-dipole.''  The polarization vector
describes asymmetries in flavor densities, such that
${\mathbf P}(p)^{initial} \propto 
\left[f_e(p,\xi_e) - f_\mu(p,\xi_\mu)\right]$,
where $f_\alpha(p,\xi_\alpha)$ is the Fermi-Dirac distribution for a neutrino
of flavor $\alpha$ with degeneracy parameter $\xi$. 

The synchronized transformation can be described by the vector equations
\begin{eqnarray}
\label{veceqns}
\partial_t{\mathbf P}_p &=& +{\mathbf A}_p \times {\mathbf P}_p + \alpha
    ({\mathbf J}- {\mathbf{\overline J}})\times {\mathbf P}_p \,, \\
\nonumber \partial_t{\mathbf{\overline P}}_p &=& -{\mathbf{\overline
    A}}_p \times {\mathbf{\overline P}}_p + \alpha ({\mathbf J}-
    {\mathbf{\overline J}})\times {\mathbf{\overline P}}_p \,,
\end{eqnarray}
where neutrino scattering is negligible, ${\mathbf J}$ denotes the individual
neutrino polarizations integrated over momentum, over-bars refer to antineutrino
quantities, and $\alpha$ is the strength of the neutrino self-potential:
$\alpha ({\mathbf J}- {\mathbf{\overline J}})\times {\mathbf P}_p$.

The general ``magnetic field'' vector ${\mathbf A}_p$ includes terms
incorporating vacuum mixing, a thermal potential from the charged-lepton
background, and a potential due to asymmetries between the charged leptons,
${\mathbf A}_p ={\vec\Delta}_p + \left[V^T(p) + V^B\right]\hat{\mathbf z}$.
Vacuum mixing is incorporated by
\begin{equation}
{\vec\Delta}_p = (\delta m_0^2/2p) (\sin 2\theta_0 {\mathbf {\hat x}} -
\cos 2 \theta_0 {\mathbf {\hat z}})\,,\label{vacuum}
\end{equation} 
where $\delta m_0^2 = m_2^2 - m_1^2$ and $\theta_0$ are the vacuum oscillation
parameters.

The thermal potential $V^T$ arises from the finite-temperature modification of
the neutrino mass due to the presence of thermally populated charged leptons,
and $V^B$ is the background potential arising due to asymmetries in charged
leptons.  $V^B$ is the crucial term in the case of the sun, but is negligible
in the early universe.

If one ignores the non-linear neutrino self-potential, the evolution of the
system is trivial: the ``magnetic-field'' vector points in the direction of the
charged lepton thermal potential, in the $\pm \hat {\mathbf z}$ direction,
which is also the initial direction of the polarization vectors in a
flavor-asymmetric system, as in DBBN.  The thermal potential initially
dominates but decreases as the universe cools, eventually becoming comparable
to $\vec\Delta_p$, the vacuum term.  $\vec\Delta_p$ points in a direction
determined by the vacuum mixing angle (Eq.~\ref{vacuum}), which for large
mixing is close to the $\hat {\mathbf x}$ direction.  Each neutrino
polarization (the flavor descriptor) then follows its respective
``magnetic-field,'' whose final orientation is in the direction of
$\vec\Delta_p$, and thus the cosmic flavor content, simply depends on the
vacuum mixing angle.

When including the neutrino self-potential, the explicit solution can only be
calculated numerically.  Ref.~\cite{abb} found that with the self-potential,
the collective system behaves on average identically with the case when the
self-potential is flatly ignored, even though the self-potential dominates all
other terms by five or more orders of magnitude.  Refs.~\cite{yyy,abb} showed
that under certain approximations, the effect of the neutrino self-potential is
to force all neutrino polarizations to follow a specific synchronization
momentum's ${\mathbf A}_p$, whose value is $ \frac{p_{\rm sync}}{T} = \pi
\sqrt{1+\xi^2/2\pi^2}\simeq \pi$, which is coincidentally very close to the
average momentum of the Fermi-Dirac distribution $\langle p/T \rangle \simeq
3.15$.  Of course, this is what the system average would follow without
self-potential.

This remarkable coincidence allows for a dramatic simplification of the
apparently initially intractable nonlinear evolution equations and allows a
straightforward visualization of the general behavior of the neutrino gas for a
variety of cases and mixing parameters.  As described above, the transformation
that leads to total or partial flavor equalization occurs at a temperature
where the vacuum term and thermal potential are comparable.  Since the vacuum
term $\vec\Delta$ is proportional to $\delta m^2$, larger $\delta m^2$ leads to
transformations at higher temperature.  And, since the final orientation of the
flavor polarization vectors is in the $\vec\Delta$ direction, the level of
total or partial flavor equalization is determined by the vacuum mixing
angle~\cite{abb}.

\section{Oscillation Parameters and The Early Universe}

The consequences of the emerging neutrino mass matrix structure for a universe
that contains neutrino degeneracies is quite rich.  The implications for each
of the mass scales in a three-neutrino mixing frame-work and their mixings is
as follows:

{\it Atmospheric Neutrinos}, $\delta m^2_{23}$ and $\theta_{23}$: for the range
of $\delta m^2$ preferred by the oscillation solution to the atmospheric
neutrino results by Super-Kamiokande~\cite{superk}, flavor equilibration occurs
at a temperature $T\sim 12\rm\, MeV$ due to the presence of equilibrating
scatterings, and maximal mixing produces absolute equalization of flavor
density asymmetries.  If precision measurements of $\theta_{23}$ reveal a
non-maximal angle, the equalization of neutrino density would be very close
though not necessarily perfect, and an explicit calculation would be necessary
since scattering is not negligible at $T\sim 12\rm\, MeV$.

{\it Solar Neutrinos}, $\delta m^2_{12}$ and $\theta_{12}$: $\delta m^2$ for
the large mixing angle solution to the solar neutrino problem is much smaller
than that of the atmospheric scale, so that the thermal potential dominates
until a lower temperature.  The transformation in this case occurs at $T\sim
2\rm\, MeV$, sufficiently before the start of nucleosynthesis at $T\sim 1\rm\,
MeV$, disallowing DBBN.  The level of equalization is dependent on the
orientation of $\vec\Delta$, i.e., how ``large'' the large mixing angle
is. Precise measurements of $\theta_{12}$ would determine the final vacuum
vector orientation, what neutrino asymmetries can be accommodated by primordial
nucleosynthesis~\cite{abb}, and therefore the maximum allowed cosmic neutrino
density.

{\it Neutrino Factories, Reactors and Long-Baseline Experiments},
$\theta_{13}$: a non-zero value $\theta_{13}$ close to the current upper limit
can lead to equalization at higher temperatures than that from the solar
scale~\cite{dhpprs}.  Also, for an inverted neutrino mass hierarchy, a very
small but non-zero $\theta_{13}$ can lead to a resonance at $T\sim 5\rm\, MeV$
that would also enhance equalization~\cite{abb}.  An appreciable $\theta_{13}$
or inverted hierarchy would further tighten the limits on the maximum cosmic
neutrino density.

In summary, the intertwining of cosmic neutrino scattering, decoupling, weak
beta-equilibrium freeze-out, and primordial nucleosynthesis with the mass and
mixing scales for neutrino transformations in degenerate cosmic neutrino
scenarios is exciting, particularly since the mass scales could have placed the
transformations much higher or lower than the primordial nucleosynthesis scale.
Therefore, the exact nature of the neutrino mass and mixing matrix, especially
if it contains further surprises, will illuminate exactly what cosmic neutrino
scenarios are plausible.


\bibliographystyle{aipproc}   

\end{document}